\documentclass[a4paper]{jpconf}

\usepackage{graphicx}

\usepackage{amsmath, amssymb, amsfonts}
\usepackage{braket}
\usepackage{color}
\usepackage{slashed}

\usepackage{textpos}
\usepackage{etoolbox}
\patchcmd{\thebibliography}{\advance\leftmargin\labelsep}
  {\labelsep=0.5cm \advance\leftmargin\labelsep}{}{}
  
\begin{document}

\begin{textblock}{4}(12,-2)
\begin{flushright}
\begin{footnotesize}
22 March 2019
\end{footnotesize}
\end{flushright}
\end{textblock}

\title{Fifth forces and discrete symmetry breaking}

\author{P Millington}

\address{School of Physics and Astronomy, University of Nottingham, \\ Nottingham NG7 2RD, UK}

\ead{p.millington@nottingham.ac.uk}

\begin{abstract}
Modifications of general relativity often involve coupling additional scalar fields to the Ricci scalar, leading to scalar-tensor theories of Brans-Dicke type. If the additional scalar fields are light, they can give rise to long-range fifth forces, which are subject to stringent constraints from local tests of gravity. In this talk, we show that Yukawa-like fifth forces only arise for the Standard Model (SM) due to a mass mixing of the additional scalar with the Higgs field, and we emphasise the pivotal role played by discrete and continuous symmetry breaking. Quite remarkably, if one assumes that sufficiently light, non-minimally coupled scalar fields exist in nature, the non-observation of fifth forces has the potential to tell us about the structure of the SM Higgs sector and the origin of its symmetry breaking. Moreover, with these observations, we argue that certain classes of scalar-tensor theories are, up to and including their dimension-four operators, equivalent to Higgs-portal theories. In this way, ultra-light dark matter models may also exhibit fifth-force phenomenology, and we consider the impact on the dynamics of disk galaxies as an example.
\end{abstract}

\section{Introduction}

Many extensions of general relativity (GR) involve additional scalar fields that are coupled non-minimally to the gravitational sector, falling into the class of scalar-tensor theories. If these additional scalar fields are sufficiently light, they can mediate long-range fifth forces. The resulting modifications of gravity are heavily constrained by local tests of gravity. However, these constraints can be avoided by appealing to screening mechanisms, wherein the fifth force is suppressed dynamically near to dense sources of matter due to the interplay between the environmental dependence and non-linearities of the governing field equations~\cite{Joyce:2014kja, Burrage:2017qrf}.

In this talk, we describe how a certain class of scalar-tensor theories, involving a coupling of the additional scalar to the Ricci scalar of GR, are equivalent to Higgs-portal theories~\cite{Silveira:1985rk, McDonald:1993ex, Burgess:2000yq, Davoudiasl:2004be, Patt:2006fw}. We show that the Yukawa-like fifth force emerges only if there is a mass mixing between the fluctuations of the new scalar $\phi$ and the would-be Higgs boson of the Standard Model (SM)~\cite{Burrage:2018dvt}. This is possible only after the electroweak phase transition and if the coupling of the new scalar breaks the discrete $\mathbb{Z}_2$ symmetry $\phi\to-\,\phi$. When these conditions are met, light Higgs portals, which might be considered as models of ultra-light dark matter in extensions of the SM, share phenomenology with modifications of gravity. We consider the impact of scalar fifth forces on the dynamics of disk galaxies as a concrete example~\cite{Gessner:1992flm, Burrage:2016yjm, OHare:2018ayv} (see also reference~\cite{Khoury:2014tka}), focussing on the symmetron screening mechanism~\cite{Hinterbichler:2010es, Hinterbichler:2011ca}, and argue that a promising direction in addressing the dark matter problem may be to exploit models that combine the phenomenology of partially screened fifth forces and ultra-light dark matter. 

\section{Coupling in new scalars}

For the SM plus GR, there are two equivalent ways of coupling new gauge-singlet scalar fields through dimension-four operators: we can either couple to the quadratic Higgs operator $H^{\dag}H$ or to the Ricci scalar $\mathcal{R}$.\footnote{In SM extensions, there may be additional operators to which we can couple the gauge-singlet scalar, e.g.~a Majorana-fermion bilinear, but the ability to modify such couplings by changing conformal frame remains.} Coupling to the former, we obtain a Higgs-portal term, e.g.
\begin{equation}
-\:\mathcal{L}\ \supset\ \beta_1\,\phi^2\,H^{\dag}H/2\;.
\end{equation}
Coupling to the latter, we instead obtain a ``Brans-Dicke term''~\cite{Brans:1961sx}, e.g.
\begin{equation}
-\:\mathcal{L}\ \supset\ \beta_2\,\phi^2\,\mathcal{R}/2\;.
\end{equation}
However, up to and including dimension-four operators, these two couplings are in fact equivalent, and they can be related to one another by a choice of conformal frame~\cite{Burrage:2018dvt}.

We start with a scalar-tensor theory of the following general form in the Jordan frame (where the non-minimal gravitational coupling is manifest):
\begin{equation}
S\ =\ \int{\rm d}^4x\;\sqrt{-\,g}\,\bigg[\frac{M_{\rm Pl}}{2}\,F(\phi)\,\mathcal{R}\:-\:\frac{1}{2}\,Z^{\mu\nu}(\phi,\partial\phi,\dots)\partial_{\mu}\phi\,\partial_{\nu}\phi-V(\phi)\bigg]\:+\:S_{\rm SM}[g_{\mu\nu}]\;.
\end{equation}
Herein, $M_{\rm Pl}$ is the reduced Planck mass and the SM fields are coupled to the Jordan-frame metric $g_{\mu\nu}$. After making a Weyl rescaling of the metric $g_{\mu\nu}\equiv F^{-1}(\phi)\tilde{g}_{\mu\nu}=A^2(\tilde{\phi})\tilde{g}_{\mu\nu}$, we can move to the Einstein frame, wherein the Einstein-Hilbert term is of canonical form, with
\begin{equation}
S\ =\ \int{\rm d}^4x\;\sqrt{-\,\tilde{g}}\,\bigg[\frac{M_{\rm Pl}}{2}\,\tilde{\mathcal{R}}\:-\:\frac{1}{2}\,\tilde{Z}^{\mu\nu}(\tilde{\phi},\partial\tilde{\phi},\dots)\partial_{\mu}\tilde{\phi}\,\partial_{\nu}\tilde{\phi}-\tilde{V}(\tilde{\phi})\bigg]\:+\:S_{\rm SM}[A^2(\tilde{\phi})\tilde{g}_{\mu\nu}]\;.
\end{equation}
When the coupling function $A^2(\tilde{\phi})$ can be expanded perturbatively, say as $A^2(\tilde{\phi})\approx 1+\tilde{\phi}^2/M^2$ --- we work throughout to leading order in terms suppressed by the scale $M$ --- the SM action can be expanded as
\begin{equation}
S_{\rm SM}[A^2(\tilde{\phi})\tilde{g}_{\mu\nu}]\ \approx\ S_{\rm SM}[\tilde{g}_{\mu\nu}]+\frac{\tilde{\phi}^2}{M^2}\,\frac{\delta S_{\rm SM}[\tilde{g}_{\mu\nu}]}{\delta \tilde{g}_{\mu\nu}}\,\tilde{g}_{\mu\nu}\ =\ S_{\rm SM}[\tilde{g}_{\mu\nu}]+\frac{1}{2}\,\frac{\tilde{\phi}^2}{M^2}\,\tilde{T}_{\rm SM}\;,
\end{equation}
where $\tilde{T}_{\rm SM}$ is the trace of the SM energy-momentum tensor. Approximating matter as a pressureless perfect fluid, $\tilde{T}_{\rm SM}= -\,A^{-1}(\tilde{\phi})\rho\approx -\,\rho$, where $\rho$ is the covariantly conserved energy density, and unit test masses feel a fifth force $\vec{F}=-\,\vec{\nabla}\ln A(\tilde{\varphi})\approx -\,\tilde{\varphi}\vec{\nabla}\tilde{\varphi}/M^2$, where $\tilde{\varphi}=\braket{\tilde{\phi}}$~\cite{Joyce:2014kja}.

However, as a result of their local scale invariance, certain terms in the matter action do not contribute to the trace of the energy-momentum tensor (on-shell); namely: gauge-field kinetic terms, fermion kinetic terms and gauge interactions, Yukawa interactions, and quartic scalar self-interactions. For the SM, this leaves the Higgs kinetic term and the (tachyonic) Higgs mass term $\mu_H^2H^{\dag}H$. The scalar kinetic term, however, gives at most derivative couplings to the non-minimally coupled scalar $\tilde{\phi}$. After redefining the SM Higgs field via $\tilde{H}\equiv A(\tilde{\phi})H$ to approach canonical normalisation, the Yukawa-like fifth force arises only if there is a mass mixing of the fluctuations $\delta\tilde{\phi}\equiv\tilde{\phi}-\tilde{\varphi}$ and those of the would-be SM Higgs boson $\delta\tilde{H}$, i.e.
\begin{equation}
-\:\mathcal{L}\ \supset\ \alpha_M\,\delta \tilde{\phi}\,\delta \tilde{H}\;.
\end{equation}
This requires that we are below the electroweak phase transition and that there is explicit or spontaneous $\mathbb{Z}_2$ breaking in the coupling function $A(\tilde{\phi})$. We then have that (in unitary gauge)
\begin{equation}
\delta\tilde{H}\ \approx\ \frac{\tilde{h}}{\sqrt{2}}\:+\:\frac{v_H}{\sqrt{2}}\,\frac{\partial A(\tilde{\varphi})}{\partial \tilde{\varphi}}\,\frac{2\mu_H^2}{m_{H}^2}\,\zeta\;,
\end{equation}
where $v_H$ is the Higgs vacuum expectation value and $m_H^2$ is (approximately) the Higgs-boson mass. Note that $2\mu_H^2/m_H^2=1$ for the SM. The light mode $\zeta$ therefore couples to the SM charged leptons as
\begin{equation}
\label{eq:leptoncoupling}
\tilde{\mathcal{L}}\ \supset\ -\:y\bar{\tilde{L}}\tilde{H}\tilde{e}_R\ \approx\ -\:(y/\sqrt{2})\bar{\tilde{e}}_L\tilde{h}\tilde{e}_R\:-\:m_e\,\frac{\partial A(\tilde{\varphi})}{\partial \tilde{\varphi}}\,\frac{2\mu_H^2}{m_{H}^2}\,\bar{\tilde{e}}_L\zeta\tilde{e}_R\;,
\end{equation}
where $m_{e}=yv_H/\sqrt{2}$ is the fermion mass and we have suppressed flavour indices.

Of course, most of the baryonic mass of the universe is due to chiral symmetry breaking. The non-minimally coupled scalar field will couple to gluons anomalously via fermion triangles, and the coupling to nucleons takes the form~\cite{Burrage:2018dvt, Gunion:1989we}
\begin{equation}
\label{eq:nucleoncoupling}
\mathcal{L}_{\rm eff}\ \supset\ -\:m_N\,\eta\,\frac{\partial A(\tilde{\varphi})}{\partial \tilde{\varphi}}\,\frac{2\mu_H^2}{m_{H}^2}\,\bar{\psi}_N\zeta \psi_N\;,
\end{equation}
where $m_N$ is the nucleon mass and $\eta$ parametrises the uncertainty in the Higgs-nucleon coupling. Interestingly, any deviation of $\eta$ from unity will lead to an effective violation of the weak equivalence principle: a positron and a proton will not fall at the same rate.

Equations~\eqref{eq:leptoncoupling} and~\eqref{eq:nucleoncoupling} illustrate two of four ways of suppressing the fifth force: (i) suppressing the explicit scale-breaking mass of the SM Higgs, i.e.~taking $\mu_H^2\to 0$, or (ii) switching off the coupling function, i.e.~driving $\partial A(\tilde{\varphi})/\partial\tilde{\varphi}\to 0$. The coupling function is suppressed dynamically in models with symmetron screening~\cite{Hinterbichler:2010es, Hinterbichler:2011ca}, which rely on symmetry breaking. The fifth force can also be suppressed (iii) by dynamically modifying the mass of the light mode $\zeta$, as in the chameleon mechanism~\cite{Khoury:2003aq, Khoury:2003rn}, or (iv) by modifying its kinetic structure, as in the Vainshtein mechanism~\cite{Vainshtein:1972sx}.
Instead, if we take $\mu_H^2\to 0$, the SM Lagrangian becomes scale invariant and the fifth force decouples~\cite{Shaposhnikov:2008xb, Brax:2014baa, Ferreira:2016kxi} (see also the Higgs-dilaton scenario~\cite{GarciaBellido:2011de}). In corollary, if light non-minimally coupled scalar fields exist in nature, the non-observation of fifth forces can be recast as an upper bound $\mu_H\lesssim 0.03\, m_H (M/M_{\rm Pl})^{1/2}$ on the explicit scale-breaking mass~\cite{Burrage:2018dvt}.

\section{Symmetron fifth forces and galactic dynamics}

\begin{figure}
\begin{center}
\includegraphics[width=0.97\textwidth]{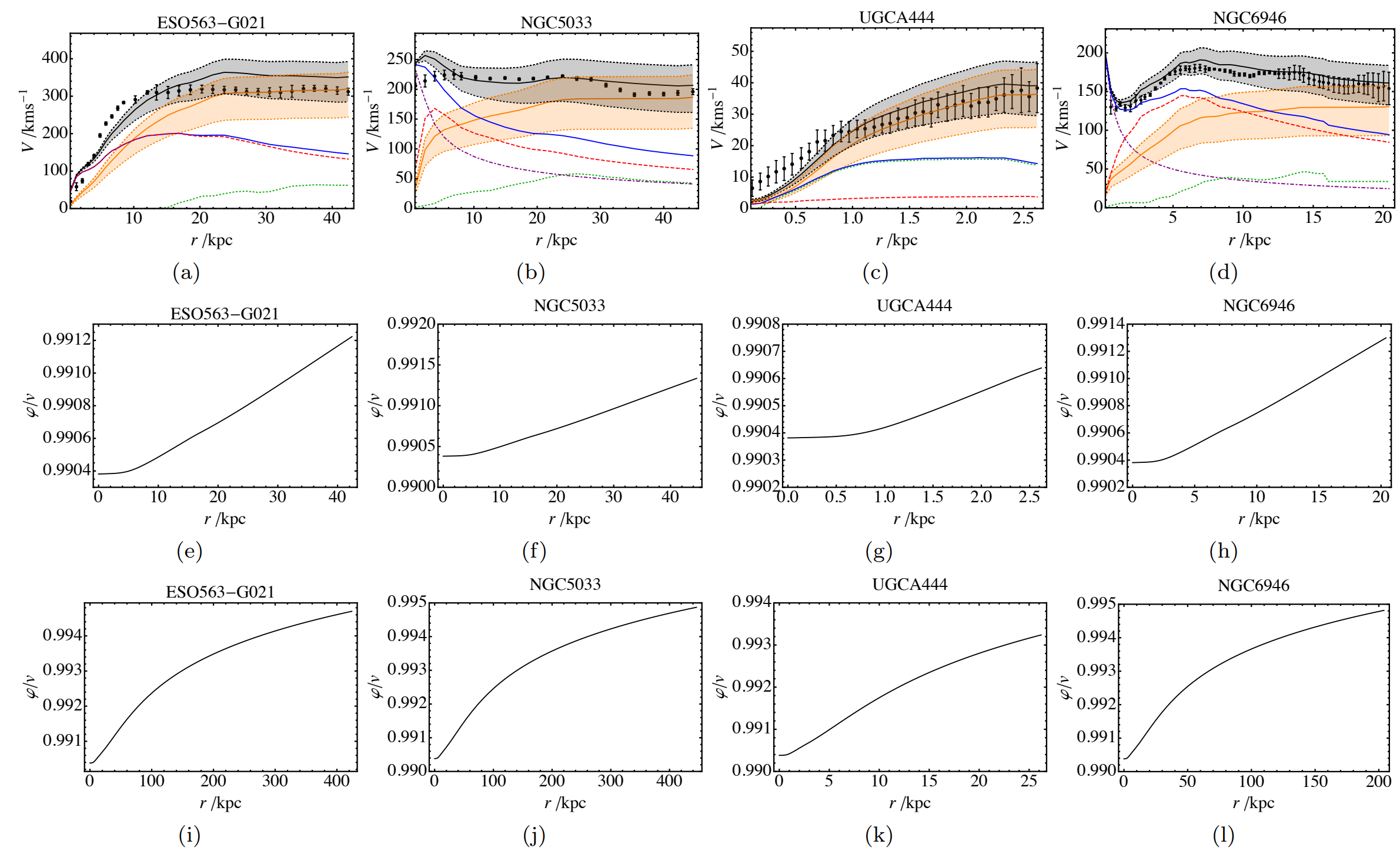}
\end{center}
\caption{\label{fig:rotationcurves}Example rotation curves (from the SPARC dataset~\cite{Lelli:2016zqa}) and field profiles ($\varphi$ in (e)--(l) is $\tilde{\varphi}$ here) for $M=M_{\rm Pl}/10$, $v/M= 1/150$ and $\mu=3\times 10^{-39}\ {\rm GeV}$. Solid black: total prediction. Solid orange: fifth-force contribution. Other lines: disk (red dashed), gas (green dotted) and bulge (purple dot-dashed) components.  Shading indicates $50\%$ variation in the coupling to the averaged baryonic density. Reprinted (figure) with permission from reference~\cite{Burrage:2016yjm}.}
\end{figure}

The prototype of the symmetron model~\cite{Hinterbichler:2010es, Hinterbichler:2011ca} (for similar models, see references~\cite{Dehnen:1992rr, Damour:1994zq, Pietroni:2005pv, Olive:2007aj, Brax:2010gi,Burrage:2016xzz}) has an effective, density-dependent potential of the form (for the canonically normalised background field $\tilde{\varphi}$)
\begin{equation}
\tilde{V}_{\rm eff}(\tilde{\varphi})\ =\ \frac{1}{2}\bigg(\frac{\rho}{M^2}-\mu^2\bigg)\tilde{\varphi}^2\:+\:\frac{\lambda}{4}\,\tilde{\varphi}^4\;.
\end{equation}
In regions of low density ($\rho<\mu^2M^2$), the $\mathbb{Z}_2$ symmetry is broken spontaneously, the symmetron field picks up a non-vanishing expectation value ($\tilde{\varphi}=v=\mu/\sqrt{\lambda}$ for $\rho=0$) and  $\partial A(\tilde{\varphi})/\partial\tilde{\varphi}\neq 0$; the fifth force is \emph{unscreened}. In regions of high density ($\rho>\mu^2M^2$), the symmetry is restored, and $\partial A(\tilde{\varphi})/\partial\tilde{\varphi}\to 0$; the fifth force is \emph{screened}. Symmetron models can also be realised by dimensional transmutation~\cite{Burrage:2016xzz}, amounting to a Higgs portal to a Coleman-Weinberg sector~\cite{Coleman:1973jx}.
  
\begin{figure}
\begin{center}
\includegraphics[width=0.97\textwidth]{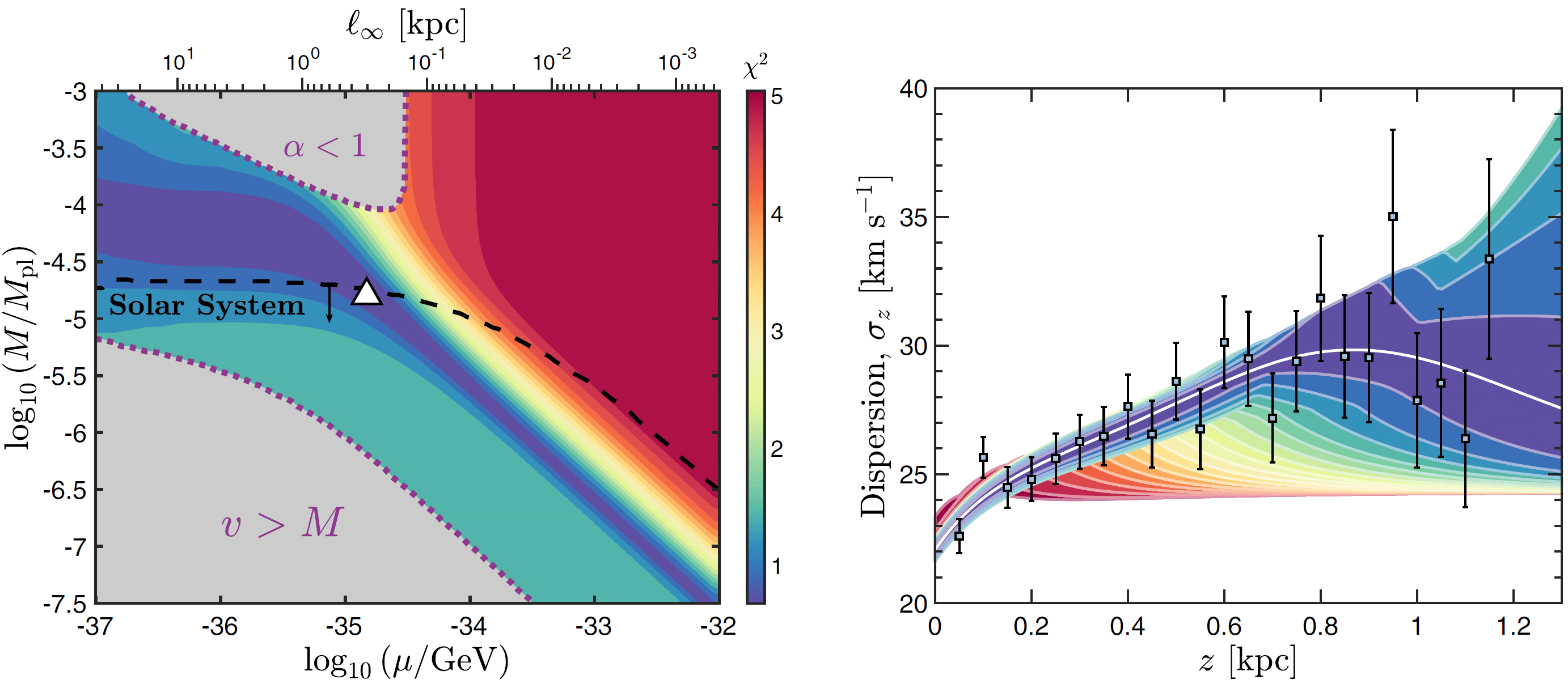}
\end{center}
\caption{\label{fig:veldisp}Left: Reduced $\chi^2$ for the fit of the symmetron fifth force model to the stellar velocity dispersion perpendicular to the galactic plane of a mock data set. Grey regions are outside the validity of the effective field theory ($v>M$) or lead to a fifth force weaker than gravity ($\alpha<1$). Right: Velocity dispersion profiles. The white line corresponds to the parameter point marked by a white triangle in the left panel. Reprinted (figure) with permission from reference~\cite{OHare:2018ayv}. Copyright (2018) by the American Physical Society.}
\end{figure}

In reference~\cite{Burrage:2016yjm}, it was shown that the symmetron profiles formed around disk galaxies can yield a fifth force that is able to explain their flat rotation curves and the correlation~\cite{McGaugh:2016leg} between the observed centripetal accelerations and those predicted from the baryonic mass alone (see figure~\ref{fig:rotationcurves} and the earlier work of reference~\cite{Gessner:1992flm}). While the benchmark parameters chosen were in conflict with Solar System tests of gravity, it was realised that there is a degenerate region of the $(\mu,M)$ parameter space along which the effective symmetron mass tends to zero in the galaxy, suggesting that it might be possible to evade Solar System constraints, while still explaining the rotation curves, by pushing deeper into the non-linear regime of the theory. This degeneracy was seen explicitly in reference~\cite{OHare:2018ayv} (see figure~\ref{fig:veldisp}), where it was shown that symmetron fifth forces might explain the observed dispersion in velocities perpendicular to the Milky Way disk. However, while these results are compelling, there are many successes of the cold dark matter paradigm that cannot be reproduced by modified gravity or for which the impact of models like the symmetron have not been studied fully. Gravitational lensing is a pertinent example.

The Maxwell term of electromagnetism is Weyl invariant, and we therefore cannot obtain a ``classical'' coupling between the scalar and photons via the conformal rescaling of the metric. Hence, if we wish to explain the discrepancy between the baryonic and lensing masses of galaxies, we must introduce additional couplings. The possibilities are reviewed in reference~\cite{Burrage:2018zuj}:

\emph{Anomalous coupling to photons:} A coupling of the scalar to photons is generated quantum mechanically via the conformal anomaly, giving rise to a dimension-six operator
\begin{equation}\vspace{-0.07em}
\mathcal{L}_{\rm eff}\ \supset\ \frac{\alpha_1}{8}\,\frac{\tilde{\phi}^2}{M_{\gamma}^2}\,\tilde{g}^{\mu\alpha}\tilde{g}^{\nu\beta}F_{\mu\nu}F_{\alpha\beta}\;.\vspace{-0.07em}
\end{equation}
In a background magnetic field, this coupling generates a mixing between the scalar and the photon mode polarised perpendicular to the magnetic field, the linear combination of which propagates on a time-like geodesic. However, the current limits of \smash{$M_{\gamma}\gtrsim 10^9\ {\rm GeV}$}~\cite{Burrage:2008ii, Burrage:2009mj, Pettinari:2010ay} (for \smash{$\alpha_1 v/M_{\gamma}\sim 1$}) preclude this mixing from inducing sufficient deflection over the GR prediction~\cite{Burrage:2018zuj}.

\emph{Photon mass:} Another option is to give the photon a mass (making it a Proca field). However, measurements of the solar wind constrain the photon mass to be less than $10^{-18}\ {\rm eV}$~\cite{Ryutov:2007zz, Tanabashi:2018oca}, and the additional deflection will always be subdominant to the GR prediction~\cite{Burrage:2018zuj}.

\emph{$\tilde{\varphi}$-dependent mass:} Alternatively, we might generate a photon mass via a term of the form
\begin{equation}
\mathcal{L}\ \supset\ \frac{\alpha_2}{4}\,\tilde{\phi}^2A_{\mu}A^{\mu}\qquad \text{or}\qquad \mathcal{L}_{\rm eff}\ \supset\ \frac{\alpha_3}{4}\,\frac{\tilde{\phi}^2}{M_{\gamma}^{\prime2}}\,(v^2-\tilde{\phi}^2)A_{\mu}A^{\mu}\;,
\end{equation}
both of which vanish in high-density environments, i.e.~when $\tilde{\varphi}=0$, and the right of which vanishes also in low-density environments, i.e.~when $\tilde{\varphi}=\pm\,v$. However, even if the spatially varying mass of the photon can avoid existing constraints, it again provides a subdominant correction to the GR prediction of the deflection angle~\cite{Burrage:2018zuj}.

\emph{Disformal coupling:} Lastly, we might introduce a coupling between the scalar and photons via a disformal transformation of the metric~\cite{Bekenstein:1992pj} (cf.~the model of reference~\cite{Khoury:2014tka})
\begin{equation}
g_{\mu\nu}\ =\ A^2(\tilde{\phi})\tilde{g}_{\mu\nu}\:+\:B(\tilde{\phi})\partial_{\mu}\tilde{\phi}\partial_{\nu}\tilde{\phi}\;.
\end{equation}
However, the scale of the disformal coupling $B(\tilde{\phi})\sim M_{\rm dis}^{-4}$ must be $M_{\rm dis}\sim 10\ {\rm meV}$ to produce an order-one change in the deflection angle~\cite{Burrage:2018zuj}, which is far below the lower bound of $M_{\rm dis}\gtrsim 650\ {\rm GeV}$ obtained from terrestrial experiments~\cite{Brax:2014vva, Brax:2015hma}.

The above constraints assume that the symmetron model is in the modified gravity regime, i.e.~the energy density of the symmetron profile does not contribute significantly to the Newtonian gravitational potential. The energy in the symmetron profile $E_{\tilde{\varphi}}$ relative to the total baryonic mass in the galaxy $M_{\rm gal}$ can be estimated to be~\cite{Burrage:2018zuj}
\begin{equation}
\label{eq:Ephi}
\frac{E_{\tilde{\varphi}}}{M_{\rm gal}}\ \approx \ \frac{\mu^2M^2}{\bar{\rho}}\,\frac{v^2}{M^2}\,I\;,
\end{equation}
where $\bar{\rho}$ is the average three-dimensional density of the galaxy and $I$ is a geometric factor (taken to be of order unity). If the galaxy is to be partially screened, so that significant fifth forces are active on galactic scales, we require $\mu^2M^2< \bar{\rho}$, in which case the energy density of the symmetron profile can be neglected. On the other hand, equation~\eqref{eq:Ephi} indicates an interesting possibility: if we can push into the regime $v/M\sim 1$ and $\mu^2M^2\sim\bar{\rho}$ then we might realise a scenario where there are contributions to the dynamics both from fifth forces and the energy density in the symmetron profile. Interestingly, the region $\mu^2M^2\sim\bar{\rho}$ coincides with the degenerate band along which the effective mass of the symmetron in the galaxy tends to zero (see figure~\ref{fig:veldisp}).

\section{Concluding remarks}

Certain classes of scalar-tensor theories are related to Higgs-portal theories via a change of conformal frame. It follows that fifth-force phenomenology depends strongly on the structure of the SM Higgs sector and the origin of its symmetry breaking, and that sufficiently light Higgs-portal theories with non-linear equations of motion may give rise to fifth forces that are subject to screening mechanisms. We have discussed the potential impact of partially screened fifth forces on galactic dynamics and concluded that the dark matter problem might be tackled by models that combine the phenomenology of ultra-light dark matter and modified gravity.

\ack

The work of PM is supported by a Leverhulme Trust Research Leadership Award. PM would like to thank his collaborators in this area: Clare Burrage, Edmund Copeland, Christian K\"{a}ding, Benjamin Elder, Ji\v{r}\'{i} Min\'{a}\v{r}, Daniela Saadeh, Michael Spannowsky and Ben Thrussell.

\end{document}